\documentclass[aps,prl,twocolumn,showpacs,superscriptaddress,groupedaddress]{revtex4}

\usepackage{amssymb}
\usepackage{color,graphicx}
\usepackage{amsmath}
\usepackage{amsbsy}
\usepackage{amsthm}
\usepackage{bbm}
\usepackage{bm}
\usepackage{epsfig}
\usepackage{lscape}
\usepackage{float}
\usepackage{subfigure}

\newcommand{\pound}{\operatornamewithlimits{\longrightarrow}}

\begin{document}

\title{Non-interferometric Test of  Collapse Models in Optomechanical Systems}

\author{M. Bahrami}
\affiliation{Department of Physics, University of Trieste, Strada Costiera 11, 34014 Trieste, Italy}
\author{M. Paternostro}
\affiliation{Centre for Theoretical Atomic, Molecular and Optical Physics, School of Mathematics and Physics, Queen\textquoteright{}s University, Belfast BT7 1NN, United Kingdom}
\author{A. Bassi}\affiliation{Department of Physics, University of Trieste, Strada Costiera 11, 34014 Trieste, Italy}
\affiliation{Istituto Nazionale di Fisica Nucleare, Sezione di Trieste, Via Valerio 2, 34127 Trieste}
\author{H. Ulbricht}
\affiliation{School of Physics and Astronomy, University of Southampton, Southampton SO17 1BJ, United Kingdom}

\date{\today}

\begin{abstract}
The test of modifications to quantum mechanics aimed at identifying the fundamental reasons behind the un-observability of quantum mechanical superpositions at the macro-scale is a crucial goal of modern quantum mechanics. Within the context of collapse models, current proposals based on interferometric techniques for their {\it falsification} are far from the experimental state-of-the-art. Here we discuss an alternative approach to the testing of quantum collapse models that, by bypassing the need for the preparation of quantum superposition states might help us addressing non-linear stochastic mechanisms such as the one at the basis of the continuous spontaneous localisation model.
\end{abstract}


\maketitle


There is clearly a growing consensus that macroscopic tests of quantum theory are one of the most promising ways to explore the boundaries between classical and quantum framework with the scope of characterising the quantum-to-classical transition. Significant theoretical and experimental efforts have been conducted so far~\cite{new_phys0,new_phys1,new_phys2,new_phys3,new_phys4,zurek}, and the interest in this area of investigation is increasing at a significant pace. 

The (so far) lack of unquestionable observations of  quantum superpositions at the macro-scale has motivated and justified the formulation of models that, by postulating an intrinsic difference between microscopic and macroscopic features, aim at pinpointing structural modifications to the Schr\"odinger equation that account for the explicit violation of the quantum superposition principle at the macroscopic level. The so-called Ghirardi-Rimini-Weber (GRW)~\cite{grw}, Continuous Spontaneous Localization (CSL)~\cite{Csl}, and Di\'osi-Penrose (DP) models~\cite{dp} are exemplary cases of the class of Collapse Models (CMs)~\cite{collapse_review1,collapse_review2} that, generally speaking, are formulated by introducing suitable stochastic non-linear terms to the Schr\"odinger equation regulating the dynamics of a quantum system. Besides embodying a key test for the quantum superposition principle, and thus a fundamental exploration of the potential limitations (if any) of the quantum framework, the experimental addressing of CMs represents a tantalising experimental challenge.

The vast majority of the proposals for the test of CMs put forward so far is based on interferometric approaches in which massive systems are prepared in large spatial quantum superposition states. In order for such tests to be effective, the superposition has to be sufficiently stable in time to allow for the performance of the necessary measurements. Needless to say, these are extremely demanding requirements from a practical viewpoint. So far, matter-wave interferometry~\cite{Hornberger2012} and cavity quantum optomechanics~\cite{aspel} are generally considered as potentially winning technological platforms in this context, and considerable efforts have been made towards the development of suited experimental configurations using levitated spheres~\cite{ori} or gas-phase metal cluster beams~\cite{Haslinger2013}.   Unfortunately, the experimental state-of-the-art is still far from allowing for a conclusive test. For instance the leading matter-wave experiment is still two orders of magnitude in mass away to test a CM~\cite{Gerlich2007, Eibenberger2013}, a challenging path as explained here~\cite{Juffmann2013}. A way forward would be the continued technical improvement of such  experimental setups, aimed at reaching suited working points. Alternatively, one might adopt a radically different approach and think of non-interferometric strategies to achieve the goal of a successful test.   

Here we explore one such possibility.
We show that CMs (in general, any nonlinear effect on quantum systems) modify the spectrum of light interacting with a radiation pressure-driven mechanical oscillator in a cavity optomechanics setting in a way that could be revealed in a simple and effective way. More specifically, we demonstrate that the CSL-affected dynamics of the mechanical oscillator results in an additional broadening term on the noise spectrum of the light driving the oscillator.  Under suitable conditions, such extra broadening can be pin-pointed to gather information on the non-linear effect due, for instance, to a collapse mechanism. By bypassing the necessity of preparing, manipulating, and sustaining the quantum superposition state of a massive object, the proposed scheme would be helpful in bringing the goal of observing CM-induced effects closer to the current experimental capabilities. 

\begin{figure}[t]%
  \includegraphics[width=0.7\columnwidth]{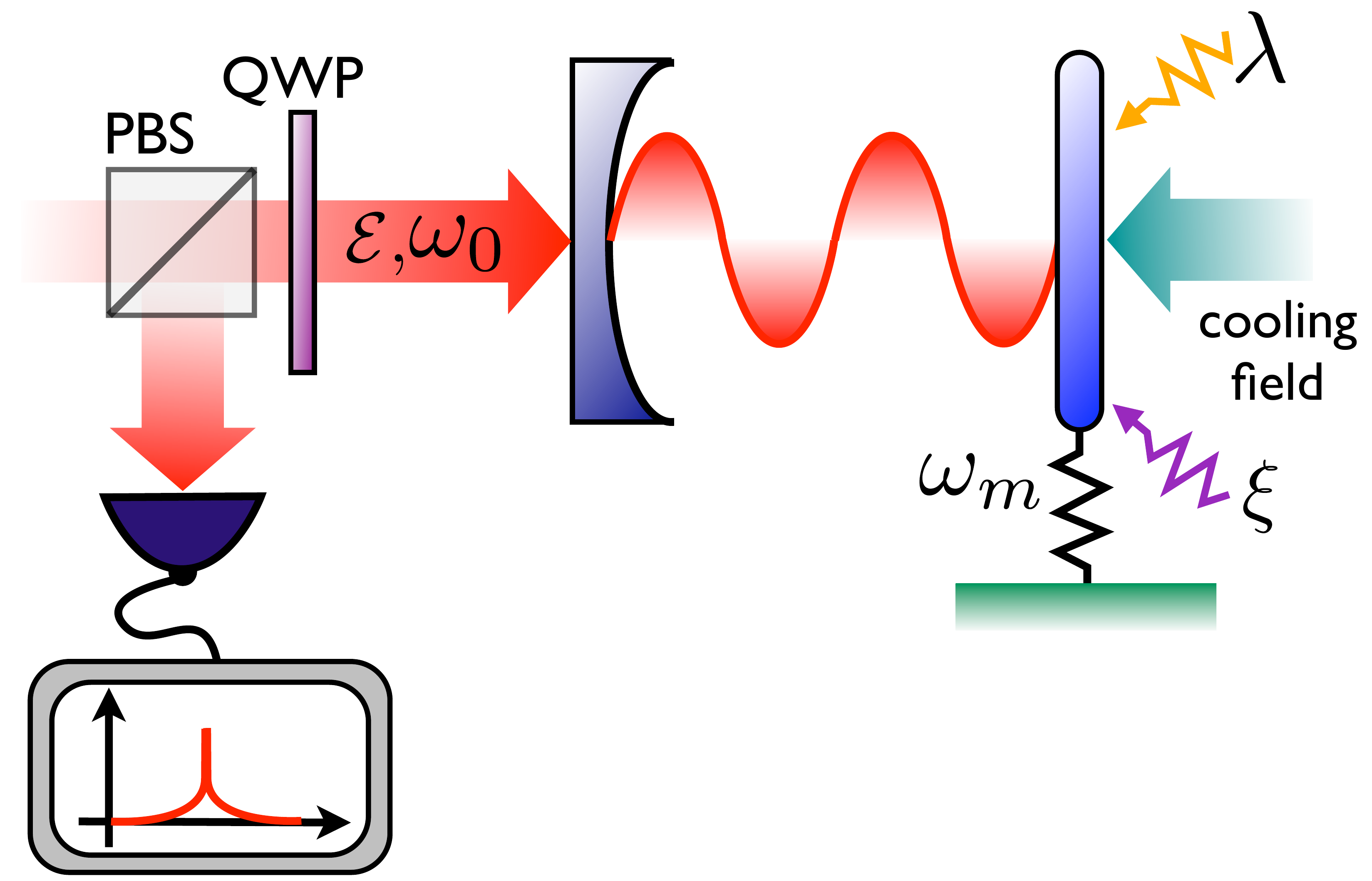}%
\caption{(Color online) Scheme of principle of the experimental setup proposed to test the CSL model. A Fabry-Perot the optomechanical cavity is pumped by a laser at frequency $\omega_0$ and strength ${\cal E}$. The pump populates a mode of the cavity filed that is coupled to a vibrating mirror (frequency $\omega_m$). A quarter-wave plate (QWP) and a polarising beam-splitter (PBS) are used to re-direct the light leaking from the cavity after the interaction with mechanical mirror, which is affected by both radiation-pressure and the non-linear mechanism responsible for CSL, to a spectrum analyser. The right-most pumping field is used to cool the mechanical oscillator to low temperatures. Zig-zag arrows are used to represent the CSL mechanism ($\lambda$) and the Brownian noise ($\xi$) affecting the mechanical oscillator.}%
 \label{scheme}%
\end{figure}%

\noindent
{\it The model.--}As anticipated, in our setting the oscillator is embodied by the moving mirror of a Fabry-Perot cavity that is driven by an external laser field. The mechanical mirror, whose oscillations are forced by its radiation-pressure coupling with the cavity field, is assumed to be in contact with a finite-temperature bath, which would in turn be  responsible for mechanical Brownian motion. In addition, we assume a non-linear mechanism to act on the oscillator, as described by a suitable CM. The setup is illustrated schematically in Fig.~\ref{scheme}. The explicitly open-system nature of the dynamics undergone by the device is fully captured by adopting a Langevin formalism to account for the Brownian noise, the leakage of the cavity field, the input white noise to the cavity, and the effect of the CM considered in our analysis. In order to set a benchmark, we concentrate on the mass-dependent Continuous Spontaneous Localization (CSL) model, which is one of the most-studied CMs in literature. The overall dynamics is thus described by the equation
\begin{equation}
\label{start}
\partial_t\hat{\mathcal{O}}=\frac{i}{\hbar}[\hat{H},\hat{\mathcal{O}}]
+\frac{i}{\hbar}[\hat{V}_t,\hat{\mathcal{O}}]+\hat{\mathcal{N}}
\end{equation}
with $\hat{\cal O}$ a generic operator of system, $\hat{H}$ the Hamitlonian relating the coherent part of the evolution, $\hat{\mathcal{N}}$ the contribution due to standard environmental noise, and $\hat{V}_t$ the instinsic noise accounted for using many-body CSL theory. 


By using Eq.~(\ref{start}) as the building block of our analysis, our goal is to show that signatures of the intrinsic collapse noise are visible in the density noise spectrum (DNS) of the mechanical oscillator. In the following, we assume the mirror to have mass $m$, natural oscillation frequency $\omega_m$, and energy damping rate $\gamma_m$. The cavity of length $L$ sustains a single mode of radiation of frequency $\omega_c$ described by the bosonic annihilation and creation operators $\hat a$ and $\hat a^\dag$. The external pump has frequency $\omega_0$ and input power $P$. In a rotating frame at the frequency of the external pump, the model Hamiltonian reads
\begin{equation}
\hat {H}=\hbar(\omega_c-\omega_0)\hat a^\dag\hat a+\frac12m\omega_m\hat q^2+\frac{\hat p^2}{2m}-\hbar\chi\hat a^\dag\hat a\hat q+i\hbar{\cal E}(\hat a^\dag-\hat a),
\end{equation}
where $\hat{q}$ is the position operator of the center-of-mass of the mechanical mirror, $\chi=\omega_c/L$ is the optomechanical coupling rate, and ${\cal E}=\sqrt{{2\kappa P}/{\hbar\omega_0}}$ quantifies the cavity-pump coupling ($\kappa$ is the cavity single-photon decay rate). The interaction term $-\hbar\chi\hat a^\dag\hat a\hat q$, which puts together the mechanical mirror and the cavity field, describes the optomechanical coupling under the assumption of large free spectral range~\cite{Law}. 
As illustrated in the Supplementary Information available at~\cite{supp}, the stochastic linear potential $\hat{V}_t $ can be cast into the form 
\begin{equation}
\label{eq:sch}
\hat{V}_t = - \hbar\,\sqrt{\lambda}\,w_t\,\hat{ q},
\end{equation}
where $w_t$ describes white noise characterized by the statistical properties $\mathbb{E}(w_t)=0$, and $\mathbb{E}(w_{t},w_{s})=\delta(t-s)$. Here $\mathbb{E}(\cdot)$ indicates expectation value and $\mathbb{E}(\cdot,\cdot)$ stands for a correlation function. Moreover~\cite{supp}
\begin{equation}
\label{lambda}
\lambda =\frac{\gamma}{3m_0^2}\sum_{k=1}^3
\int\,
\frac{e^{-\frac{|\mathbf{r}-\mathbf{r}'|^2}{4r_C^2}}}{(2\sqrt{\pi}\,r_C)^3}
\partial_{r_k}\varrho({\bf r})
\,\partial_{r'_k}\varrho({\bf r}')d{\bf r}\,d\mathbf{r}'\,
\end{equation}
with $m_0=1\,$amu, $\varrho({\bf r})$ the mass density of the mechanical mirror, $r_C=10^{-7}$m a characteristic length entering the CSL model, and $\gamma$ a coefficient that measures the strength of the coupling with collapse noise. 
Ghirardi, Pearle and Rimini~\cite{Csl} set  $\gamma_{\text{\tiny GRW}} \simeq 10^{-36}\text{m}^{3}\text{s}^{-1}$, while Adler~\cite{adlerphoto} sets $\gamma_{\text{\tiny A}} \simeq 10^{-28}\text{m}^{3}\text{s}^{-1}$. Much larger or smaller values are ruled out~\cite{Csl,adlerphoto}. As a benchmark for the quantification of $\lambda$, one can consider a homogeneous spherical object of radius $R$ and mass $m$. Using Eq.~(\ref{lambda}), one thus gets
\begin{equation}
\lambda\approx\frac{3\gamma\,m^2}{8\pi^{\frac32}m_0^2r_C\,R^4}(1-e^{-R^2/r_C^2})~~~[m_0=1\,\text{amu}].
\end{equation}

Let us now get back to Eq.~(\ref{start}). We now have all the ingredients to write explicitly as a set of quantum Langevin equations reading~\cite{mauro}  
\begin{equation}
\begin{aligned}
\partial_t\hat{q}&=\hat{p}/m,\\
\partial_t\hat{p}&=-m\omega_m^2\hat{q}+\hbar\,\chi\,\hat a^\dag \hat a-\gamma_m\,\hat{p}+\hat{\xi}+\hbar\sqrt{\lambda}\,w_t,\\
\partial_t\hat{a}&=i(\omega_0-\omega_c)\hat{a}+i\,\chi\,\hat{q}\,\hat{a}-\kappa\,\hat{a}+\sqrt{2\kappa}\,\hat{a}_{in}.
\end{aligned}
\end{equation}
where 
we have introduced the 
cavity input noise operator $\hat a_{in}$, the Brownian-motion Langevin operator $\hat\xi$ (describing the incoherent motion of the mechanical mirror arising from the coupling with the background of phononic modes due to its physical support). These sources of noise are characterized by the two-time correlators~\cite{mauro}
\begin{equation}
\label{corre}
\begin{aligned}
&\mathbb{E}(\hat \xi(t),\hat \xi(t'))=\frac{\hbar m\gamma_m}{2\pi}\int\omega e^{-i\omega(t-t')}[\coth(\beta\omega)+1]d\omega,\\
&\mathbb{E}(\hat a_{in}(t),\hat{a}_{in}(t'))=
\mathbb{E}(\hat a^\dag_{in}(t),\hat{a}_{in}(t'))=0,\\
&\mathbb{E}(\hat a_{in}(t)\hat{a}^\dag_{in}(t'))=\delta(t-t')
\end{aligned}
\end{equation}
with $\beta=\hbar/(2k_BT)$, $k_B$ the Boltzmann constant, and $T$ the temperature of the phononic bath with which the mechanical mirror is at equilibrium. This set of equations is in general very difficult to solve due to the non-linear nature of the optomechanical coupling (see very recent progress towards the treatment of the full non-linear process in Ref.~\cite{Simon}). However, under the assumption of large pumping (i.e. large input power of the driving field), we can expand the field and mirror operators in fluctuations around their respective mean values $\overline{\bm v}$ as $\hat{\bm v}=\overline{\bm v}+\delta\hat{\bm v}$ with ${\bm v}=(q,p,a)$. The steady-state mean values can be easily determined and used to derive a simplified set of equations for the fluctuation operators that can be solved in frequency space~\cite{mauro}. Leaving the details of an otherwise straightforward calculation aside, we can focus on the form of the symmetrized two-frequency correlation function $S(\omega)\delta(\omega+\Omega)=\mathbb{E}(\hat q(\omega)\hat q(\Omega)+\hat q(\Omega)\hat q(\omega))/2$, which embodies the DNS of the mirror's position. By assuming Markovianity of the mechanical Brownian motion, which justified in the limit of moderate temperature and small mechanical damping, we get
\begin{widetext}
\begin{equation}
\label{eq:S_w}
S(\omega)=\frac{2\alpha_s^2 \hbar^2 \kappa  \chi ^2 (\Delta ^2+\kappa^2 + \omega^2 )+\hbar m \omega
    [(\Delta ^2+\kappa^2-\omega^2)^2 +4\kappa^2\omega^2][{\gamma_m} \coth\left(\beta\omega\right)+\Lambda]}{\left|2\alpha_s^2 \Delta  \hbar \chi ^2+m \left(\omega ^2-\omega_m^2-i\gamma_m \omega \right) \left[\Delta ^2+(\kappa +i \omega)^2\right]\right|^2}
\end{equation}
\end{widetext}
with $\Lambda=\lambda\,(\hbar/m\omega_m)$, $\Delta\simeq{\omega_c-\omega_0}$ the cavity-pump detuning, and $\alpha_s={\cal E}/\sqrt{\kappa^2+\Delta^2}$ being the steady-state amplitude of the cavity field. Eq.~(\ref{eq:S_w}) is the key formal result of this analysis and the focus of the analysis that we will present in the remainder of this work. It is worth stressing that an alternative approach to the calculations presented here would be the explicit modification of the two-time correlator $\mathbb{E}(\hat \xi(t),\hat \xi(t'))$ in Eq.~\eqref{corre} with the replacement $\hat\xi\to\hat\xi+\hbar\sqrt{\lambda}\omega_t$ and the Markov approximation for the mechanical Brownian motion.

\begin{figure*}[t!]
{\bf (a)}\hskip7.5cm{\bf (b)}
\includegraphics[width=1.1\columnwidth]{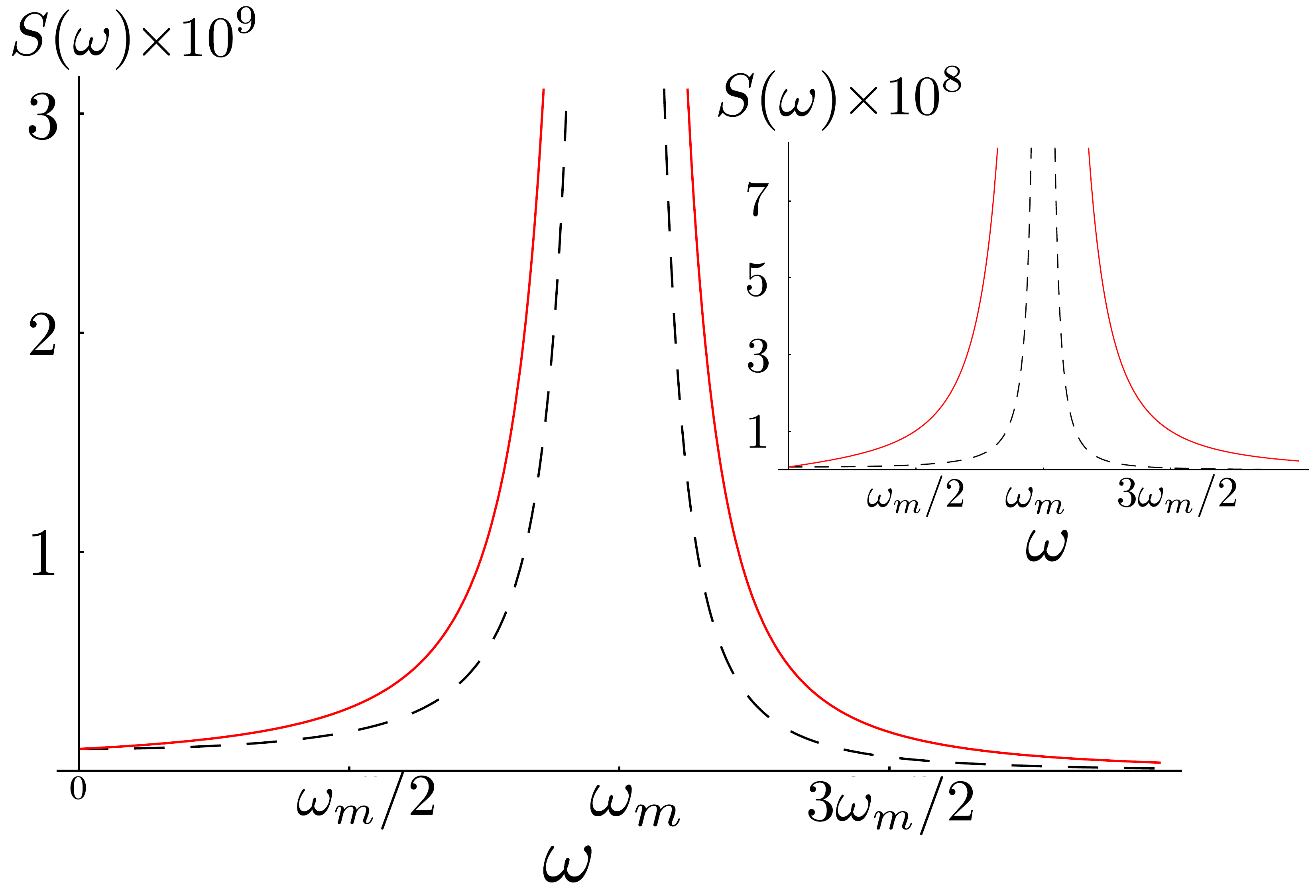}~~~~\includegraphics[width=0.9\columnwidth]{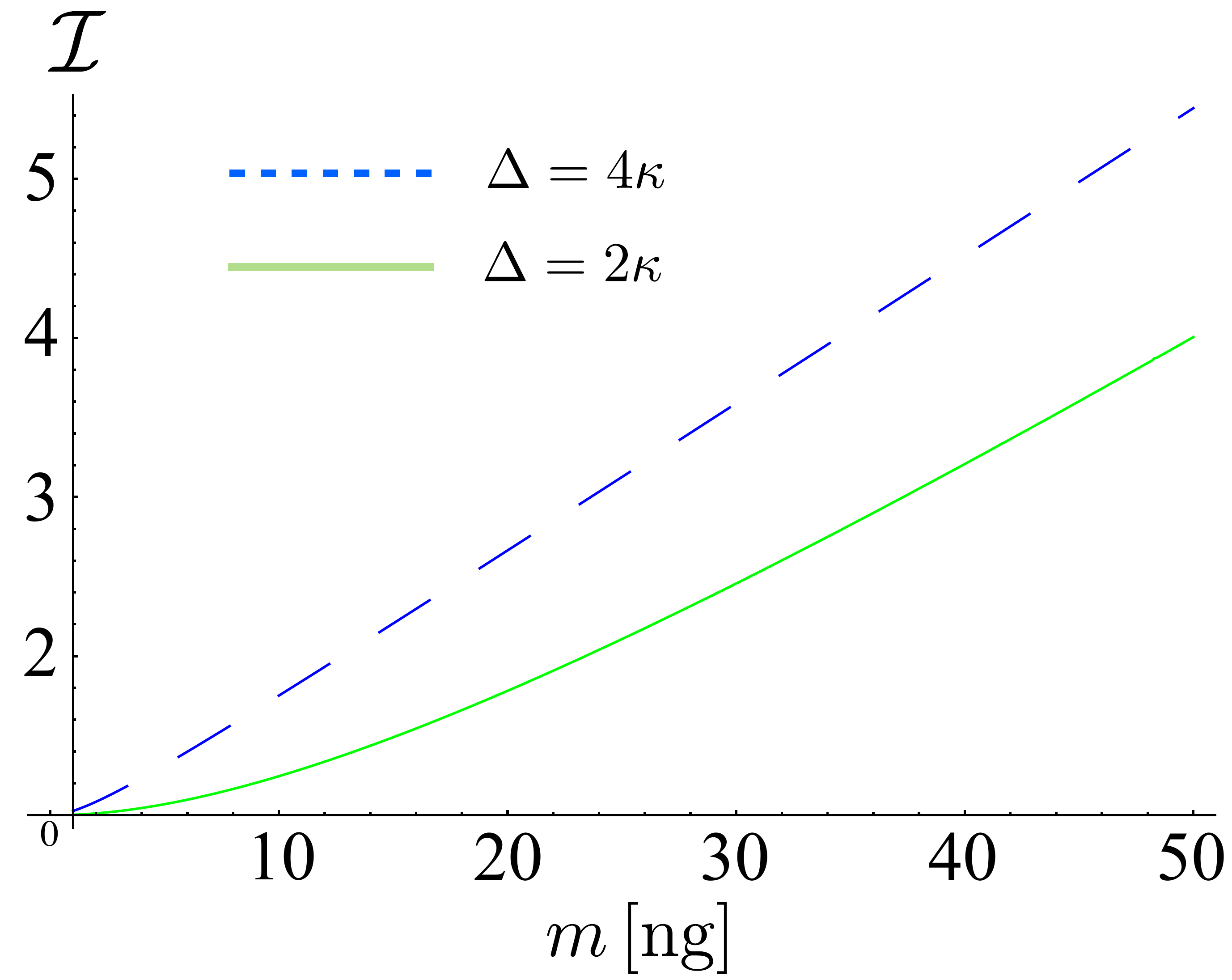}
\caption{(Color online) {\bf (a)} Main panel: DNS $S(\omega)$ against the frequency $\omega$ for $\omega_m/2\pi=2.75\times10^5$Hz, $\gamma_m/2\pi=\omega_m/10^5$, 
$L=25$mm, ${\cal P}=4$mW, $\kappa=5\times10^7$Hz, $T=1$mK, and for a cantilever of $1\mu$m of linear dimension. We have compared the DNS without any CSL effect (black dashed curve) to the one corresponding to $\lambda=\lambda_{A}$ (red solid line), for $m=15$ng. Inset: Same as the main panel but for $m=150$ng. All curves are evaluated at $\Delta=4\kappa$, which results in a lower effective temperature of the mechanical mirror. {\bf (b)} We plot the area underneath the DNS $S(\omega)$ against $m$ for two choices of the detuning and $\lambda=\lambda_A$. Other parameters are as in Fig.~\ref{DNS}{\bf (a)}. The case of no CSL mechanism corresponds to a horizontal line at ${\cal I}=1$.}
\label{DNS}
\end{figure*}

\noindent
{\it Discussion.--}
Clearly, the CSL mechanism manifests itself in the DNS as an addition to the thermal contribution embodied by $\coth(\beta\omega)$ [cf. Eq.\eqref{eq:S_w}]. It is thus immediate to realise that in order to magnify the effect due to the CSL up to the point of making it observable, we should deplete the thermal contributions to the DNS. This requires a low initial temperature of the mechanical mirror and, possibly, the use of an additional radiation pressure-based passive cooling mechanisms that brings the mechanical system in equilibrium at a lower temperature than that of its surrounding bath. This is equivalent to assuming that the mirror is in a thermal state at a low effective temperature $T$. Moreover, by arranging for a large detuning $\Delta$, we can achieve conditions such that the alleged CSL mechanisms is actually key in determining the steady-state conditions of the mechanical mirror. In fact, from the expression above, and using the definitions of the parameters entering $S(\omega)$, one can see that 
\begin{equation}
\lim_{(\beta,\Delta)\to\infty}S(\omega)\simeq\frac{\hbar(\gamma_m+\Lambda)}{m\omega_m\gamma_m^2}\pound_{\gamma_m\ll\Lambda}\frac{\hbar\Lambda}{m\omega_m\gamma_m^2}.
\end{equation}
Needless to say, the situation is not achievable in actual experiments, where only a finite detuning and a non-zero temperature are achievable in practice. However, as we will see, this dos not preclude, in principle, the observable nature of the CSL effects.  

In Fig.~\ref{DNS} {\bf (a)}, for instance, we compare the DNS of the mechanical mirror with and without CSL effects at a moderately large detuning and for values of the key parameters that are not far from experimental realizability. Clearly, by acting like an additional term to the natural thermal broadening of the noise spectrum of the mirror, the CSL mechanisms results in a wider $S(\omega)$ and does not affect the peak position of the spectrum. This suggests that an effective way to determine it quantitatively would be to calculate the area underneath the spectrum, thus inferring the modification that such additional term induces on the average energy of the mirror. We have thus considered the quantity
\begin{equation}
{\cal I}=\frac{\int^\infty_{-\infty}S(\omega)\,d\omega}{\int^\infty_{-\infty}S_{\Lambda\to0}(\omega)\,d\omega},
\end{equation}
which gives us a quantitative estimate of the relative increase of the area under the DNS for $\Lambda\neq0$ with respect to the case of no CSL mechanism. 
In Fig.~\ref{DNS} {\bf (b)} we show the behavior of such a figure of merit against the value of $\Lambda$, calculated using the parameter $\gamma_{\rm A}$ (which is much larger than the estimate provided by Ghirardi, Pearle and Rimini, and thus offers more chances of being actually observed), and for some choices of the detuning: at large detunings, ${\cal I}$ appears to be a linear function of $\Lambda$ and for $\Delta/\kappa\gtrsim 1$ (we notice that we work in the bad-cavity regime, which is much easier to achieve, experimentally), and the CSL mechanism could result in a sizeable increase of the DNS area. Needless to say, the actual value by which such area increases strongly depends on the set of parameters that are used to model the dynamics of the system and by no means we claim for optimality. It is worth mentioning, finally, that as shown in Ref.~\cite{mauro} and experimentally demonstrated in many optomechanics experiments, the light leaking out of the cavity can be used to reconstruct the spectrum of the intra-cavity mechanical mirror. The use of standard input-output relations $\delta{\hat a}_{out}+\delta{\hat a}_{in}=\sqrt{2\kappa}\delta{\hat a}$, linking the extra-cavity field to the input noise and the intra-cavity signal, shows that the same signatures of the CSL mechanism persist in the extra-cavity signal, which can be effectively used to infer the value of $\Lambda$ and, from this, the characteristic parameter $\lambda$ of the CSL model. Details of this analysis are provided in Ref.~\cite{supp}.

\noindent
{\it Conclusions.}-- Our analysis supports the idea that the effects of non-linear stochastic modifications to quantum mechanics, such as those that characterize the CSL model, are observable by adopting an indirect approach that does not rely on the ad hoc creation of a quantum superposition state. We have illustrated such a possibility using a cavity optomechanics setting where the noise properties of the field leaking from a Fabry-Perot cavity with a vibrating end mirror carry information on potential influences due to collapse-like mechanisms. Our proposal appears to only require low operating temperatures of the mechanical mirror, a condition that is met in most of the cutting-edge experiments in cavity optomechanics reported so far~\cite{aspel}. Moreover, although we have focused our discussion to the end-mirror configuration of an optomechanical setup, it is clearly perfectly suited to be adapted to both membrane-in-the-middle and levitating-nanosphere configurations, therefore embodying a general paradigm of vast appeal. 

\noindent
{\it Acknowledgements.---} MP thanks the UK EPSRC for a Career Acceleration Fellowship and a grant awarded under the ``New Directions for Research Leaders" initiative (EP/G004579/1), and the John Templeton Foundation (grant 43467).
MB and AB acknowledge financial support from the EU project NANOQUESTFIT. 
AB wish to thank the COST Action MP1006 ``Fundamental Problems in Quantum
Physics" and acknowledges partial support from INFN. HU is supported by EPSRC (EP/J014664/1), the Foundational Questions Institute (FQXi), and the John Templeton foundation (grant 39530).

\begin{widetext}

\section*{Supplementary Information}

\section{Formal analysis of the model}

In this Section, we show how the inclusion of an intrinsic collapse field that interacts with a many-body system affects the center-of-mass motion of the system itself, as prescribed in Collapse Models.
We shall focus on the Continuous Spontaneous Localization (CSL) collapse model, which is the most studied collapse model in the literature. Taking the system as a rigid body and averaging over the relative coordinates, we derive the corrections to the Schr\"odinger equation due to the CSL collapse field for cases where the spread of the wavefunction of the center-of-mass in smaller than the CSL correlation length $r_C=10^{-7}\,$m.

Violation of quantum superposition is phenomenologically described by nonlinear equations replacing Schr\"odinger?s equation.
Collapse models determine the class of nonlinear modi?cations of the Schr\"odinger equation which are compatible with the assumption of no-faster-than-light signalling~\cite{signal4}. Since with violations of the superposition principle we mean superpositions in space, therefore the evolution of the wave function reads:
\begin{equation}
\label{eq:wave_x}
\begin{aligned}
\frac{d}{dt}|\Psi_t\rangle &=
\left[
-\frac{i}{\hbar}\hat{H}_0
+\sqrt{\gamma}\,\int\,d\mathbf{x}\,\left(\hat{L}({\bf x})-\langle \hat{L}({\bf x}) \rangle_{\Psi_t}\right)\,\xi_t({\bf x})
-\frac{\gamma}{2}\,\int\,d\mathbf{x}\,\left(\hat{L}({\bf x})-\langle \hat{L}({\bf x}) \rangle_{\Psi_t}\right)^2
\right]|\Psi_t\rangle
\end{aligned}
\end{equation}
where $\xi_t({\bf x})$ is a noise-field, white both in space and time, and 
$\langle \hat{L}({\bf x}) \rangle_{\Psi_t}=\langle \Psi_t |\hat{L}({\bf x})|\Psi_t \rangle$ which induces the nonlinearity in the dynamics. The statistical properties of the noise field and the form of Lindblad operator $\hat{L}({\bf x})$ is determined by the model.

The many-body Lindblad operator of the mass-proportional Continuous Spontaneous Localization (CSL) model, the most-studied collapse model in the literature~\cite{Csl,collapse_review1,collapse_review2}, is given by
\begin{equation}
\label{eq:noise-csl}
\hat{L}({\bf x})=
\int \,d\mathbf{y}\,g(\mathbf{x-y})\sum_{\text{j}}\frac{m_\text{j}}{m_0}\sum_s\hat{a}_{\text{j}}^{\dagger}(s,\mathbf{y})\hat{a}_{\text{j}}\left(s,\mathbf{y}\right),
\end{equation} 
where $m_0=1\,$amu, $\gamma \simeq 10^{-28}\,\text{m}^{3}\text{s}^{-1}$, $\hat{a}_{\text{j}}\left(s,\mathbf{y}\right)$ is the annihilation operator of particle of type-$\text{j}$ with mass $m_\text{j}$ and the spin $s$ at position $\mathbf{y}$; and
$g({\bf r}) = \exp(-\mathbf{r}^{2}/2r_{C}^{2})/(\sqrt{2\pi}r_{C})^{3}$
with $r_C \simeq 10^{-7}\,$m the correlation length.
In fact, in the CSL model a system is well-localized when its position spread is smaller than $r_C$. For the CSL noise field, which is a universal noise, we have $\mathbb{E}(\xi_t(\mathbf{x}),\xi_{\tau}(\mathbf{y}))=\delta(t-\tau)\,\delta(\mathbf{x}-\mathbf{y})$, with $\mathbb{E}(\cdots)$ the stochastic average.

Since we work in non-relativistic regime of quantum theory, in Eq.~\eqref{eq:noise-csl} the type of particles runs over electrons and nucleons where the number of particles is also a constant of motion. Accordingly, in the subspace of a fixed number of particles, we can write:
$
\hat{L}(\mathbf{x})\approx
\sum_{j=1}^{N}\,A_j\,g(\mathbf{x}-\hat{\mathbf{x}}_{j})
$, 
where $N$ is the number of atomic nuclei, $A_j$ is the atomic mass number, and $\mathbf{x}_{j}$ is the nuclear position. 
Separating the center-of-mass motion from relative ones for a rigid system, the Lindblad operator reads:
\begin{eqnarray}
\label{eq:Lx_r}
\hat{L}(\mathbf{x})\simeq\frac{1}{m_0}
\,\int \,d\mathbf{r}\,\varrho(\mathbf{r})\,g(\mathbf{x}-{\mathbf{r}}-\hat{\mathbf{q}}).
\end{eqnarray}
with $\hat{\bf q}$ the center-of-mass position operator and $\varrho(\mathbf{r})$ the mass density of the system.  

For the case where the center-of-mass is distributed around a time-dependent mean value $\langle \hat{\mathbf{q}}\rangle_{\Psi_t}$ with the spread much smaller than $r_C$, one can Taylor-expand $g(\mathbf{x}-{\mathbf{r}}-\hat{\mathbf{q}})$ to the first order around $\langle \hat{\mathbf{q}}\rangle_{\Psi_t}$. After somehow lengthy (yet straightfroward) calculations, one finally obtains
\begin{eqnarray}
\label{eq:wave_q}
\frac{d}{dt}|\psi_t(\mathbf{q})\rangle &=&
\left[
-\frac{i}{\hbar}\hat{H}_\mathbf{q}
+\sqrt{\gamma}\,\left(\hat{{\bf q}}-\langle \hat{{\bf q}} \rangle_{\Psi_t}\right)
\cdot\tilde{\mathbf{w}}_t
-\frac{\gamma\,\eta}{2}
\left(\hat{{\bf q}}-\langle \hat{{\bf q}}\rangle_{\Psi_t}\right)^2
\right]|\psi_t({\bf q})\rangle,
\end{eqnarray}
where $\hat{H}_\mathbf{q}$ is the standard quantum Hamiltonian of the center-of-mass. Moreover, we have
\begin{eqnarray}
\label{eq:eta}
\eta &=&\frac{1}{m_0^2}\sum_{k=1}^3
\int\,d{\bf r}\,d\mathbf{r}'\,\varrho(\mathbf{r})\,\varrho(\mathbf{r}')\,
\frac{e^{-\frac{|\mathbf{r}-\mathbf{r}'|^2}{4r_C^2}}}{(2\sqrt{\pi}\,r_C)^3}
\left(\frac{1}{2r_C^2}-\frac{(r_k-r'_k)^2}{4r_C^4}\right)\\
&=&\frac{1}{m_0^2}\sum_{k=1}^3
\int\,d{\bf r}\,d\mathbf{r}'\,
\frac{\exp\left[-\frac{|\mathbf{r}-\mathbf{r}'|^2}{4r_C^2}\right]}{(2\sqrt{\pi}\,r_C)^3}
\partial_{r_k}\varrho({\bf r})
\,\partial_{r'_k}\varrho({\bf r}'),
\end{eqnarray}
with ${\bf r}=(r_1,r_2,r_3)$, 
and $\tilde{\mathbf{w}}_t$ a three-dimensional white-noise term
\begin{equation}
\tilde{\mathbf{w}}_t=\frac{1}{m_0}\int\, d\mathbf{x}\,\xi_t(\mathbf{x})\,\int \,d\mathbf{r}
\,g(\mathbf{x}-{\mathbf{r}}-\langle\hat{\mathbf{q}}\rangle_{\Psi_t})\,
\partial_{{\bf r}}\varrho({\bf r})
\end{equation}
that is characterised by the correlation function
\begin{equation}
\mathbb{E}(\tilde{w}_{k,t},\tilde{w}_{l,s})=
\,\delta(t-s)\,\delta_{kl}\,\frac{1}{m_0^2}\,
\int d{\bf r}\,d\mathbf{r}'\,
\frac{e^{-\frac{|\mathbf{r}-\mathbf{r}'|^2}{4r_C^2}}}{(2\sqrt{\pi}\,r_C)^3}\partial_{r_k}\varrho({\bf r})
\,\partial_{r'_k}\varrho({\bf r}').
\end{equation}
Here, $k,l=1,2,3$ indicate the spatial components of $\tilde{\mathbf{w}}_t$.

Instead of attacking directly the nonlinear dynamics given in Eq.\eqref{eq:wave_q}, we address the Schr\"odinger equation with a stochastic potential
\begin{equation}
\label{eq:sch}
i\hbar \frac{d}{dt} |\psi_t({\bf q})\rangle =(\hat{H}_{\bf q} + \hat{V}_t) |\psi_t({\bf q})\rangle,~~~~ \hat{V}_t = - \hbar\,\sqrt{\gamma}\,\hat{{\bf q}}\cdot \tilde{{\bf w}}_t.
\end{equation}
As often discussed in literature~\cite{stoch1,stoch2,stoch3,stoch4,stoch5}, the effects of nonlinear terms in Eq.\eqref{eq:wave_q}, at the statistical level, can be mimicked also by linear random potentials. For individual realizations of the noise, the consequences of such different approaches are rather distinct. At the statistical level, however, they coincide if the potential is suitably chosen.
For one-dimensional cases, it is possible to recast 
Eq.~\eqref{eq:sch} as 
\begin{equation}
i\hbar \frac{d}{dt} |\psi_t(q)\rangle =(\hat{H}_{q} + \hat{V}_t) |\psi_t(q)\rangle,~~~~ \hat{V}_t = - \hbar\,\sqrt{\lambda}\,\hat{q} \, w_t.
\end{equation}
where $w_t$ describes white noise with $\mathbb{E}(w_t)=0$ and $\mathbb{E}(w_t\,w_s)=\delta(t-s)$, and
\begin{equation}
\lambda=\frac{\gamma}{3m_0^2}\,\sum_{k=1}^3\,
\int d{\bf r}\,d\mathbf{r}'\,
\frac{e^{-\frac{|\mathbf{r}-\mathbf{r}'|^2}{4r_C^2}}}{(2\sqrt{\pi}\,r_C)^3}\partial_{r_k}\varrho({\bf r})
\,\partial_{r'_k}\varrho({\bf r}').
\end{equation}

\section{Analysis of the extra-cavity field}

In this Section we provide an analysis of the field leaking out of the optomechanical cavity proposed in the main Letter and describe in some details the formal steps needed in order to reconstruct the density noise spectrum (DNS) of the intra-cavity signal, thus inferring the effect of the CSL mechanism at the core of our investigation. 

In the main Letter we provide all the necessary ingredients to determine the DNS of the intra-cavity field [cf. Eq.~(8) and related discussions]. As hinted there and discussed in details in Ref.~\cite{mauro}, this passes through the determination of the steady-state solutions of a set of dynamical equations for the fluctuation operators of the optomechanical system that are derived, straightforwardly, from Eq.~(6) in the main Letter. By going to the frequency domain and introducing the fluctuation of the cavity field's quadrature operators $\delta\hat x=\delta\hat a+\delta\hat a^\dag$ and $\delta\hat y=i(\delta\hat a^\dag - \delta\hat a)$ (and the analogous quantities for the input noise operators), we get
\begin{equation}
\begin{aligned}
i\omega\delta\hat x(\omega)&=\Delta\delta\hat y(\omega)+\sqrt{2\kappa}\delta\hat{x}_{in}(\omega)-\kappa\delta\hat x(\omega),\\
-i\omega\delta\hat y(\omega)&=-\Delta\delta\hat x(\omega)+\sqrt{2\kappa}\delta\hat y_{in}(\omega)\kappa\delta\hat y(\omega)+2\alpha_s\chi\delta\hat q(\omega),\\
-i\omega\delta\hat q(\omega)&=\delta\hat p(\omega)/m,\\
-i\omega\delta\hat p(\omega)&=-m\omega^2_m\delta\hat q(\omega)+\hbar\chi\alpha_s\delta\hat q(\omega)-\gamma\delta\hat p(\omega)+\hat\xi(\omega)+\hbar\sqrt{\lambda}\tilde w_\omega,
\end{aligned}
\end{equation}
where $\tilde w_\omega$ is the Fourier transform of variable $w_t$ and $\delta\hat v(\omega)$ ($v=x,y,q,p,x_{in},y_{in}$) stands for the frequency-domain version of the fluctuation operator $\delta\hat v$. By solving for $\delta\hat y(\omega)$, using the input-output relation $\delta\hat y_{out}+\delta\hat y_{in}=\sqrt{2\kappa}\delta\hat y$~\cite{inout}, and calculating the DNS for $\delta\hat y_{out}$, we find
\begin{equation}
\label{out}
S_{y_{out}}(\omega)\simeq1+\frac{4\alpha^2_s\chi^2(\kappa^2+\omega^2)}{|\Delta^2+(\kappa-i\omega)^2|^2}S(\omega)
\end{equation}
with $S(\omega)$ the DNS of the mirror's position reported in Eq. (8) of the main Letter. Some inessential terms have been left out of Eq.~\eqref{out} with only negligible differences with respect to the exact result~\cite{mauro}. The unit term in Eq.~\eqref{out} entails the contribution coming from the input shot noise. The proportionality between $S_{y_{out}}(\omega)$ and $S(\omega)$, albeit non trivial, allows us to reconstruct the CSL effects on the area underneath the DNS of the outgoing field quadrature's fluctuations $\delta\hat y_{out}$. In Fig.~\ref{outfigure} we show the results corresponding to the same situation addressed, for the intra-cavity field, in Fig. 2 {\bf (b)}. Clearly enough, the CSL effect is virtually unaltered as far as the extra-cavity DNS is concerned.

\begin{figure}[t!]
\includegraphics[width=0.7\columnwidth]{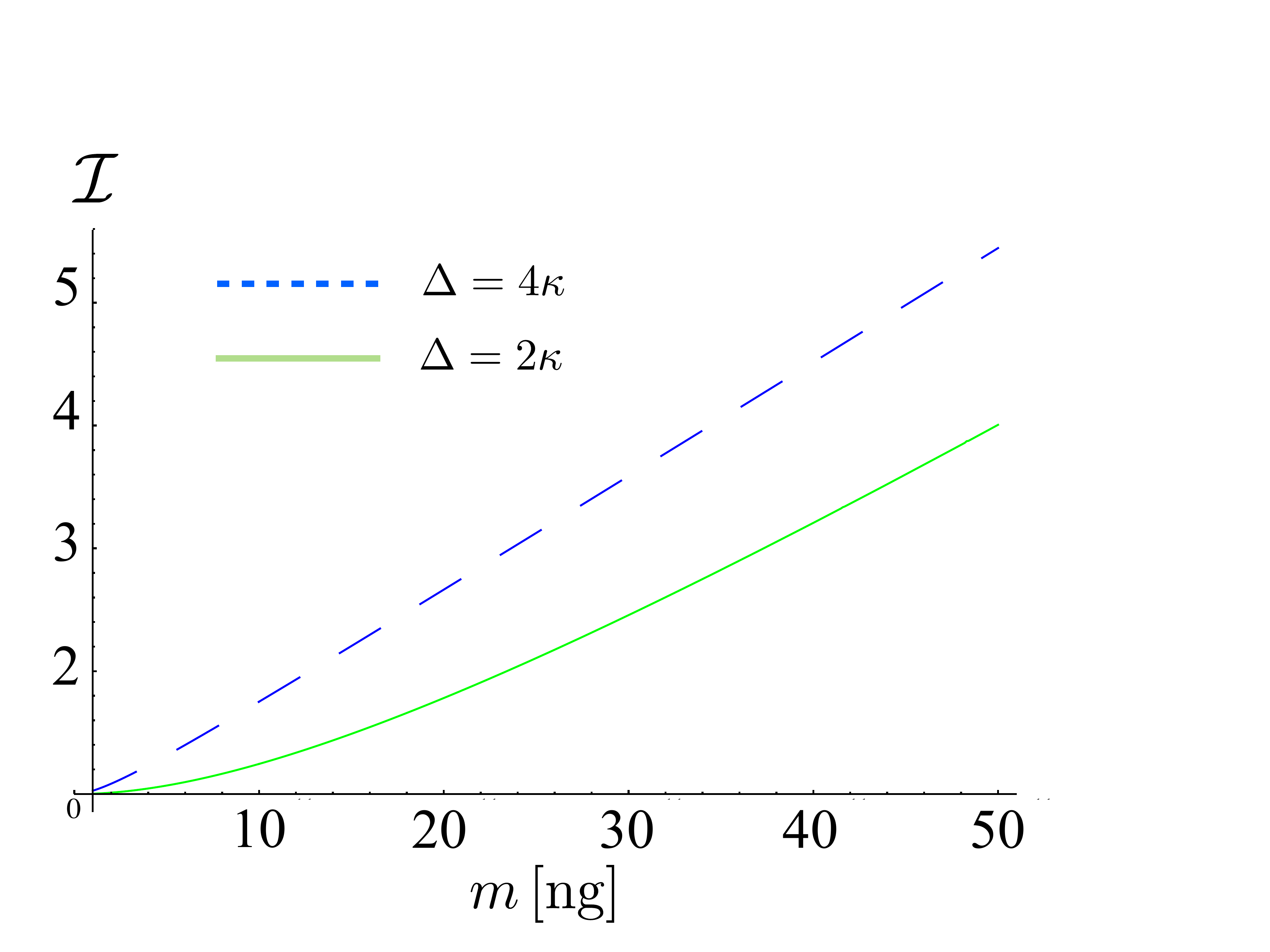}
\caption{(Color online) Area underneath the DNS of the operator $\delta\hat y_{out}$ (stripped of the shot-noise contribution) for the same parameters used in Fig. 2 {\bf (b)} of the main Letter. }
\label{outfigure}
\end{figure}

\end{widetext}



\begin{thebibliography}{99}
\providecommand{\url}[1]{\texttt{#1}}
\providecommand{\urlprefix}{URL }
\providecommand{\eprint}[2][]{\url{#2}}


\bibitem{new_phys0} 
A. J. Leggett, 
J. Phys.: Condens. Matter \textbf{14}, R415 (2002).

\bibitem{new_phys1} 
S. L. Adler, {\it Quantum Theory as an Emergent Phenomenon} (Cambridge University Press, 2004).

\bibitem{new_phys2} 
S. L. Adler, A. Bassi, 
Science {\bf 325}, 275 (2009).

\bibitem{new_phys3} S. Weinberg, 
Phys. Rev. A \textbf{85}, 062116 (2012).

\bibitem{new_phys4} R. Penrose, Gen. Relativ. Gravit. {\bf 28}, 581 (1996).

\bibitem{zurek} W. H. Zurek, arXiv:quant-ph/0306072.

\bibitem{grw}
G.C. Ghirardi, A. Rimini, and T. Weber, Phys. Rev. D {\bf 34}, 470 (1986).

\bibitem{Csl}
G. C. Ghirardi, P. Pearle, A. and Rimini, Phys. Rev. A {\bf 42}, 78 (1990);
G. C. Ghirardi, R. Grassi, and F. Benatti, Found. Phys. {\bf 25}, 5 (1995).

\bibitem{adlerphoto}
S. L. Adler, J. Phys. A {\bf 40}, 2935 (2007).

\bibitem{dp}
L. Di\'{o}si, Phys. Lett. A {\bf 120}, 377 (1987);
R. Penrose, Gen. Rel. Grav. \textbf{28}, 581 (1996).

\bibitem{collapse_review1}
A. Bassi, and G. C. Ghirardi, Phys. Rep. {\bf 379}, 257 (2003).

\bibitem{collapse_review2}
A. Bassi, K. Lochan, S. Satin, T. P. Singh, and H. Ulbricht, Rev. Mod. Phys. {\bf 85}, 471 (2013).

\bibitem{Hornberger2012} K. Hornberger, S. Gerlich, P. Haslinger, S. Nimmrichter, and M. Arndt, Rev. Mod. Phys. {\bf 84}, 157 (2012).

\bibitem{aspel}  F. Marquardt and S. M. Girvin, Physics {\bf 2}, 40 (1993); M. Aspelmeyer, S. Gr\"oblacher, K. Hammerer, and N. Kiesel, J. Opt. Soc. Am. B {\bf 27}, A189 (2010); M. Aspelmeyer, T. J. Kippenberg, and F. Marquardt, arXiv:1303.0733 (2013).

\bibitem{ori} O. Romero-Isart, A. C. Pflanzer, F. Blaser, R. Kaltenbaek, N. Kiesel, M. Aspelmeyer, and J. I. Cirac, Phys. Rev. Lett. {\bf 107}, 020405 (2011); 
O. Romero-Isart, Phys. Rev. A {\bf 84}, 052121 (2011); B. Pepper, R. Ghobadi, E. Jeffrey, C. Simon, and D. Bouwmeester, Phys. Rev. Lett. {\bf  109}, 023601 (2012); J. Bateman, S. Nimmrichter, K. Hornberger, and H. Ulbricht, arXiv:1312.0500 (2013).

\bibitem{Haslinger2013} P. Haslinger, N. D\"orre, P. Geyer, J. Rodewald, S. Nimmrichter, and M. Arndt, Nat. Phys. {\bf 9}, 144 (2013).


\bibitem{Gerlich2007} S. Gerlich, L. Hackerm\"uller, K. Hornberger, A. Stibor, H. Ulbricht, F. Goldfarb, T. Savas, M. M\"uri, M. Mayor, and M. Arndt, Nat. Phys. {\bf 3}, 711 (2007).

\bibitem{Eibenberger2013} S. Eibenberger, S. Gerlich, M. Arndt, M. Mayor, and J. T\"uxen, Phys. Chem. Chem. Phys. {\bf 15}, 14696 (2013).

\bibitem{Juffmann2013} T. Juffmann, H. Ulbricht, and M. Arndt, Rep. Prog. Phys. {\bf 76}, 086402 (2013).

\bibitem{supp} See supplementary material for more detail.

\bibitem{mauro} 
M. Paternostro, S. Gigan, M. S. Kim, F. Blaser, H. R. B\"ohm and M. Aspelmeyer, {New J. Phys.} {\bf 8}, 107 (2006). 

\bibitem{Law} C. K. Law, Phys. Rev. A {\bf 49}, 433 (1994); C. K. Law, Phys. Rev. A {\bf 51}, 2537 (1995).

\bibitem{Simon} B. He, Q. Lin, R. Ghobadi, and Ch. Simon, arXiv:1308.5932 (2013).

\bibitem{signal4}
A. Bassi, D. D\"urr, G. Hinrichs, Phys. Rev. Lett. {\bf 111}, 210401 (2013).

\bibitem{stoch1}
N. Gisin, 
Hel. Phys. Acta {\bf 62}, 363 (1989);
Phys. Lett. A {\bf 143}, 1 (1990).

\bibitem{stoch2}
N. Gisin, and M. Rigo, 
J. Phys. A {\bf 28}, 7375 (1995).

\bibitem{stoch3}
J. Polcinski, 
Phys. Rev. Lett. {\bf 66}, 397 (1991).

\bibitem{stoch4} S. L. Adler, J. Math. Phys. \textbf{41}, 2485 (2000); hep-th/0206120 (2002), Section 5F; Phys. Rev. D \textbf{67}, 025007 (2003), Added note.

\bibitem{stoch5} L. Di\'{o}si, L. Phys. Lett. A \textbf{129}, 419 (1988). 

\bibitem{inout} Gardiner C W 1991 Quantum Noise (Berlin: Springer); M. J. Collett, and C. W. Gardiner, Phys. Rev. A {\bf 30}, 1386 (1984).

\end{thebibliography}
\end{document}